\documentclass[%
 reprint,
 amsmath,amssymb,
 aps,
]{revtex4-2}

\usepackage{graphicx}
\usepackage{dcolumn}
\usepackage{bm}
\usepackage{hyperref}
\usepackage{upgreek}

\begin{document}

\preprint{APS/123-QED}

\title{Low-noise Balanced Homodyne Detection with Superconducting Nanowire Single-Photon Detectors}

\author{Maximilian Protte}
\thanks{These authors contributed equally to this work\\$^\dagger$ timon.schapeler@upb.de}
\author{Timon Schapeler$^\dagger$}
\thanks{These authors contributed equally to this work\\$^\dagger$ timon.schapeler@upb.de}
\author{Jan Sperling}
\author{Tim J. Bartley}
\affiliation{Institute for Photonic Quantum Systems, Department of Physics, Paderborn University, Warburger Str. 100, 33098 Paderborn, Germany}

\date{\today}

\begin{abstract}
Superconducting nanowire single-photon detectors (SNSPDs) have been widely used to study the discrete nature of quantum states of light in the form of photon-counting experiments. We show that SNSPDs can also be used to study continuous variables of optical quantum states by performing homodyne detection at a bandwidth of $400~\mathrm{kHz}$. By measuring the interference of a continuous-wave field of a local oscillator with the field of the vacuum state using two SNSPDs, we show that the variance of the difference in count rates is linearly proportional to the photon flux of the local oscillator over almost five orders of magnitude. The resulting shot-noise clearance of $(46.0\pm1.1)~\mathrm{dB}$ is the highest reported clearance for a balanced optical homodyne detector, demonstrating their potential for measuring highly squeezed states in the continuous-wave regime. In addition, we measured a $\mathrm{CMRR}=22.4~\mathrm{dB}$. From the joint click counting statistics, we also measure the phase-dependent quadrature of a weak coherent state to demonstrate our device's functionality as a homodyne detector.
\end{abstract}

\maketitle

\section{Introduction}
Balanced homodyne detection (BHD) is a method frequently used in quantum optics to study the wave-like nature of optical quantum states in terms of their continuous variables~\cite{Lvovsky2009,Andersen2015}. A pertinent example of continuous variables comprise the quadrature representation of the optical quantum state, as opposed to the discrete-variable photon-number representation. The wave-like nature of the field results from the coherence between different photon number components of a field with uncertain photon number. Therefore, to study the wave-like nature of the field, one must be able to measure superpositions of photon number states.

To do so, the field of a weak optical quantum state (signal) comprising a few photons interferes with the field of a bright coherent state (i.e., a local oscillator - LO) on a balanced beam splitter. The two output modes of the beam splitter are then measured with two photodetectors. By calculating the difference between the detector count rates, which is proportional to the field quadrature at a reference phase provided by the local oscillator, the optical quantum state can be characterized~\cite{Yuen1983, Schumaker1984, Braunstein1990}. The ability to characterize optical quantum states in their phase-space representation~\cite{Lvovsky2009} makes homodyne detection an essential tool for quantum information processing with continuous variables~\cite{Braunstein2005}. 

Typically, conventional semiconductor photodiodes are used as the detector in balanced homodyne detection. The high optical flux arising from the LO lifts the optical signal above the electronic noise floor, which is typically in the $\textrm{pW}\sqrt{\textrm{Hz}}$ range at telecommunication wavelengths. As a result, the generated carriers in the photodiode can be integrated over a characteristic response time, resulting in a photocurrent which is proportional to the incident optical flux~\cite{Yariv1991}. For BHD, it is essential that the photodetector output is directly proportional to the intensity (photon flux) of the input light. In this case, we refer to the detector output as linear.

By contrast, superconducting nanowire single-photon detectors (SNSPDs) show extremely low noise down to the $10^{-9}~\textrm{pW}\sqrt{\textrm{Hz}}$ level~\cite{Chiles2022}. Combined with their high efficiency~\cite{Reddy2020,Chang2021} and excellent timing resolution~\cite{Korzh2020}, the question arises whether they can be configured for homodyne detection. At first glance, SNSPDs seem unsuitable for homodyne detection due to their highly nonlinear response to the input photon number distribution. Because of the on-off characteristic of SNSPDs, we consider them as nonlinearly proportional to the input photon flux since they cannot distinguish between one or more photons impinging on the detector. However, in certain regimes of photon flux, such detectors can be used for phase-dependent measurements~\cite{Banaszek1996}. Here the field to be measured is interfered with a strong LO on a highly transmissive beam splitter, generating a displacement of the field in quadrature space proportional to the LO's amplitude. Kovalyuk et al. demonstrated a coherent detection with SNSPDs in a heterodyne nanophotonic device~\cite{Kovalyuk2017a}. Alternatively, weak-field homodyne detection (WFHD) is a method whereby phase-dependent photon statistics following interference with a weak local-oscillator are measured. This has been demonstrated before with avalanche photodiodes~\cite{Donati2014} and transition edge sensors~\cite{Thekkadath2020}. This method can be used to violate a Bell inequality~\cite{Kuzmich2000}. However, one major disadvantage of WFHD is the sensitivity to noise. In conventional homodyne detection, the bright local oscillator acts as a mode filter since only modes matched to the LO are optically amplified and measured in BHD. When using on/off detectors and a weak LO, the signal field that does not overlap with the LO mode will still contribute detection events to the measurement, leading to noise.

In this paper, we show that with SNSPDs and a strong LO, we can perform a mode-selective BHD. Under continuous-wave illumination, the rate at which an SNSPD registers a detection event is randomly distributed in time. The click rate is linearly proportional to the input photon flux when the incident flux results in a detection event rate exceeding the detector noise floor. However, this linear relationship is limited to a finite photon flux, above which the detector saturates.

To demonstrate this, we perform a homodyne measurement at a bandwidth of $400~\mathrm{kHz}$ of the vacuum state under varying local oscillator photon fluxes. We show that the variance of the difference count rate of two SNSPDs is linearly proportional to the input photon flux over a wide range of input fluxes. From this we infer that the detector count rates are linearly proportional to the input photon flux in this range. We show that the difference count rates from the SNSPDs is shot-noise limited and exhibits a shot-noise clearance of $(46.0\pm1.1)~\mathrm{dB}$, which is the highest reported clearance for an optical homodyne detector~\cite{Jin2015, Bruynsteen2021}. We then perform a homodyne measurement of a weak coherent state with a photon flux of $\approx4.3~\mathrm{kphotons/s}$ and calculate from the joint click counting statistics the phase-dependent quadrature of a weak coherent state.

\section{Concept and Experimental Setup}
\begin{figure}[t!]
\centering\includegraphics[scale=1]{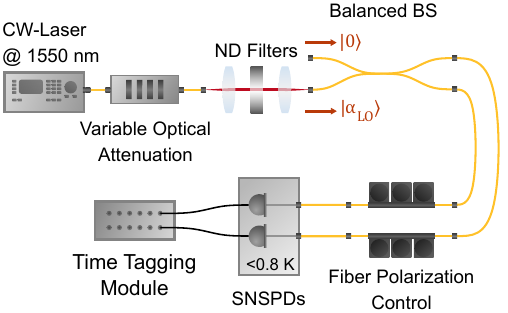}
\caption{Setup used to investigate the linearity of two SNSPDs in terms of the local oscillator photon flux. The field of the vacuum state is interfered with a continuous-wave (CW) local oscillator on a balanced fiber beam splitter. The splitting ratio of the fiber coupled beam splitter was measured independently with a power meter to be $(51\pm1.8)\%$ and $(49\pm1.8)\%$ for each arm. The variable optical attenuation is used to investigate the variance of the difference count rate as a function of the local oscillator photon flux. Additional neutral-density filters are used to attenuate the local oscillator photon flux, in order to make it compatible with the SNSPDs. The polarization control can be used to level the count rates measured on the SNSPDs for the subtraction.}
\label{fig:1}
\end{figure}

To investigate the linear operation regime of SNSPDs in the context of homodyne detection, we use the setup shown in Fig.~\ref{fig:1}.
A fiber-coupled continuous-wave laser source at $1550~\mathrm{nm}$ with a spectral bandwidth of $400~\mathrm{kHz}$ corresponding to a coherence time of $2.5~\mathrm{\upmu s}$, is used as the local oscillator. The photon flux of the local oscillator is controlled using a variable optical attenuator (VOA). The VOA is varied from $1-86~\mathrm{dB}$, resulting in a set of local oscillator photon fluxes varying from $\approx 0.07~\mathrm{photons/s}$ to $\approx2.3\cdot10^7~\mathrm{photons/s}$ ($9\times10^{-21}~\mathrm{W}-3\times10^{-12}~\mathrm{W}$).

Below the minimal attenuation (i.e. maximum photon flux) set on the VOA, the detectors latch. In this case, the detectors do not return to the superconducting state and are insensitive to subsequent incident photons. At the highest attenuation, the count rate is given by the dark counts of the detectors. Therefore, we can evaluate the detectors over their complete operation range. We interfere the field of the local oscillator and the vacuum state on a balanced fiber beam splitter and couple the light into a cryostat with two commercial single-pixel SNSPDs (Photon Spot). The SNSPD efficiencies were measured to be $(85\pm5)\%$ and $(92\pm5)\%$. The two SNSPDs are biased to maintain high efficiency at low dark counts. Since the two SNSPDs have unequal efficiency, the polarization controller in front of the detectors can be used to fine-tune the relative efficiency of the SNSPDs and balance the measured count rates on the SNSPDs. We calculate the overall efficiency after coupling back to fiber to be $(77\pm7)\%$. This includes the transmission through the beam splitter, and the efficiency of the SNSPDs. The transmission is measured with calibrated power meters.
\begin{figure*}[!ht]
\centering\includegraphics[scale=1]{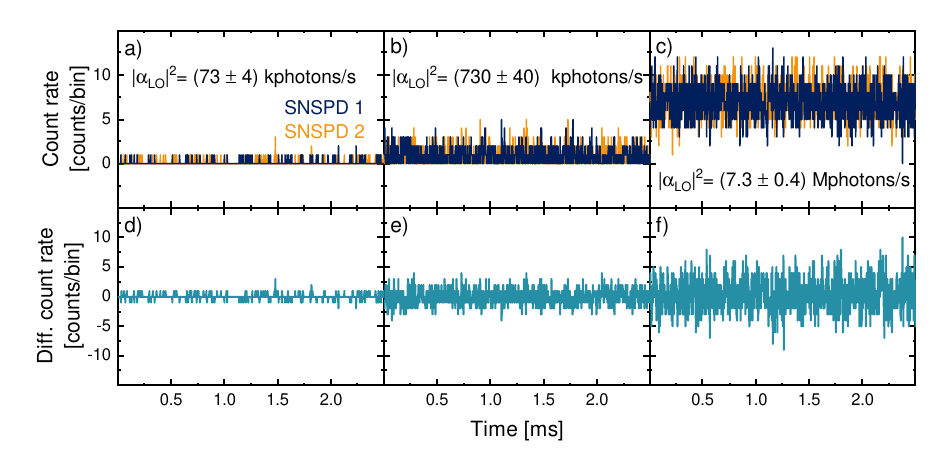}
\caption{a)-c) count rate measured on both SNSPDs in units of photons per bin. The bin size was $2.5~\mathrm{\upmu s}$. The count rate is shown for three different local oscillator photon flux values of a) $(73\pm4)~\mathrm{kphotons/s}$, b) $(730\pm40)~\mathrm{kphotons/s}$ and c) $(7.3\pm0.4)~\mathrm{Mphotons/s}$. From the count rates, we calculate the difference as shown in d)-f). An increasing noise of the difference count rate is visible for increasing local oscillator photon fluxes.}
\label{fig:2}
\end{figure*}

A time-tagging module records the detection events at each detector independently. We set an artificial dead time of the time tagger channel to $100~\mathrm{ns}$, which prevents the occurrence of additional false counts due to triggering on noise on the falling edge of the SNSPD's electrical signal. This dead time limits the sampling rate of the homodyne detector to $10~\mathrm{MHz}$. The BHD is operated with a sampling rate of $400~\mathrm{kHz}$, corresponding to the bandwidth of the LO, ensuring an operation of the BHD in a single temporal mode. For higher sampling rates, the measurement time would be reduced, coming at the cost of a reduced resolution with which the quadrature can be measured. With faster detectors, the concept could be extended to spectral broader light sources.

The dead time of SNSPDs considerably limits the operation window for homodyne detection. For the linear operation, the probability of more than one photon arriving at the detector within the dead time must be low. Therefore, characterization under continuous-wave illumination yields the maximum LO photon flux that the detectors can handle since the photons are maximally distributed over time. For pulsed optical signals, the maximum LO photon flux is thereby reduced, leading to reduced shot-noise clearance. Utilizing detectors with a smaller dead time would allow for higher photon flux.

To later measure the quadrature and variance with the highest possible resolution , we choose a time bin of $2.5~\mathrm{\upmu s}$, matching the bandwidth of the LO. Within a time bin, which is inversely proportional to the sampling rate of the BHD, the measured counts are summed. The counts within a time bin are acquired over a duration of $5~\mathrm{s}$. The resulting count rate is shown in Fig.~\ref{fig:2} for three different local oscillator photon flux values of (a) $(73\pm4)~\mathrm{kphotons/s}$, (b) $(730\pm40)~\mathrm{kphotons/s}$ and (c) $(7.3\pm0.4)~\mathrm{Mphotons/s}$. The count rates show the balanced operation of the two SNSPDs for different local oscillator photon flux values. 

To show the linearity of the SNSPD count rate with respect to the incident photon flux, we calculated the difference from the two SNSPD count rates and calculate the variance of the difference count rate. The variance of the difference is given by~\cite{Gerry2004}
\begin{equation}
	\mathrm{Var}(\hat{n}_{\mathrm{diff}})=4|\beta|^2\mathrm{Var}(\hat{X}_\varphi)\,,
 \label{eq:1}
\end{equation}
where $|\beta|^2$ is the intensity of the local oscillator in terms of the photon flux and $\mathrm{Var}(\hat{X}_\varphi)$ is the variance of the phase-dependent quadrature. $\varphi$ is defined as the optical phase between the field of the LO and the optical quantum state. $\hat{n}_{\mathrm{diff}}$, for which the variance is calculated, is defined as the difference in the count rates measured at the two SNSPDs. Even if the experiments are performed at small photon fluxes, considering the average photon flux within a sampling window, we can assume that the experiments are performed in the strong LO limit, with a more than 100-fold difference between the photon flux of the LO and the signal state~\cite{Schumaker1984}.

Since we will investigate the vacuum state, for which $\mathrm{Var}(\hat{X}_\varphi)=1/4$, we expect that $\mathrm{Var}(\hat{n}_{\mathrm{diff}})_\mathrm{vac}\propto|\beta|^2$, from which we see that with a linear increase of the local oscillator photon flux a linear increase of the measured variance is expected. This boundary given by the local oscillator photon flux is referred to as the shot-noise limit. In Fig.~\ref{fig:2} (d)-(f) the difference count rate calculated from the count rates in Fig.~\ref{fig:2} (a)-(c) is shown. From the difference count rate, a clear scaling of the noise with the local oscillator photon flux is visible. For the following investigation we first check if this scaling is linear with respect to the input photon flux.

\section{Homodyne Detector Characterization}
\begin{figure}[!ht]
\centering\includegraphics[scale=1]{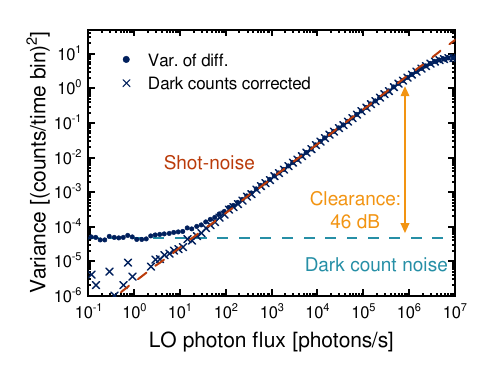}
\caption{Variance of the measured difference count rate as a function of the local oscillator photon flux. For a low local oscillator photon flux, the variance is dominated by the variance of the dark count rate. An increase in local oscillator photon flux leads to a linear increase of the variance. The measured variance fits well to the shot-noise limit with a slope of $1$. When fitting a linear regression to the variance of the difference count rate, the gradient of the linear region is given by $0.996\pm0.002$. At high photon flux, the variance drops below the shot noise, indicating a non-linear behavior of the SNSPDs. We measure a shot-noise clearance of $(46.0\pm1.1)~\mathrm{dB}$ for a detector sampling rate of $400~\mathrm{kHz}$ and $2~\mathrm{MSamples}$.}
\label{fig:3}
\end{figure}

In order to investigate how the variance scales with the local oscillator photon flux, we scanned the entire range from $0.07-2.3\cdot10^7~\mathrm{photons/s}$. We measured the counts on each SNSPD for a measurement time of $5~\mathrm{s}$ with a time bin size of $2.5~\mathrm{\upmu s}$, which gives a detector sampling rate of $400~\mathrm{kHz}$. This results in $2~\mathrm{MSamples}$ over which we calculate the variance. The resulting variance is shown in Fig.~\ref{fig:3}. The errors for the local oscillator photon flux are given by the $5\%$ uncertainty of the power meter. The error bars in the variance, which scale with $[1/(2n-2)]^{1/2}$ where $n$ is the number of samples used for the calculation~\cite{Squires2001}, are not visible.

For a low local oscillator photon flux, a constant variance is measured, which results from the dark count rate measured on both SNSPDs. The dark count variance $V_\mathrm{dark~counts}=(18\pm2)~\left[\mathrm{counts/s}\right]^2$ (dashed blue line in Fig.~\ref{fig:3}) is averaged over the data points for which the variance is constant and corresponds to the variance in the difference count rate for zero local oscillator power. With increasing local oscillator photon flux, an increase in the variance is observable. The dashed red line indicates the shot-noise limit which is given by $\mathrm{Var}(\hat{n}_{\mathrm{diff}})_\mathrm{vac}\propto|\beta|^2$, where we calculate the local oscillator photon flux $|\beta|^2$ from the measured local oscillator power, the losses in the setup and the additional attenuation.From Fig.~\ref{fig:3}, it is evident that with increasing local oscillator photon flux, the variance is well-described by the shot-noise limit and that the SNSPD count rate can be treated as linear in a limited photon flux regime.

At even higher local oscillator photon fluxes ($>1\cdot10^6~\mathrm{photons/s}$), a clear deviation from the linear regime is visible. At this point the probability of a second photon arriving at a detector within the dead time of the detector after a detection event is not negligible. As a result, the local oscillator photon flux in units of $\mathrm{photons/s}$ is not properly measured. By subtracting the dark count variance from the measured data, we can extend the linear regime down to the noise floor dictated by the dark counts.

To analyze the performance of our homodyne detector further, we investigate the shot-noise clearance from the data shown in Fig.~\ref{fig:3}. The shot-noise clearance is given by
\begin{equation}
    \mathrm{Clearance}=-10\log\left(\frac{V_\mathrm{dark~counts}}{V_\mathrm{lin,max}}\right)[\mathrm{dB}] \,,
 \label{eq:2}
\end{equation}
where $V_\mathrm{dark~counts}$ is the variance in the measured difference count rates for zero local oscillator power (i.e. the variance of the dark counts) and $V_\mathrm{lin,max}$ is the variance in the measured difference count rates for the maximum local oscillator photon flux, where the detectors are still considered linear. To define the upper limit of our detector, we considered a maximum deviation of $0.1$ of the calculated variance from the shot-noise limit. Thereby, we consider the homodyne detector count rate as linear up to a local oscillator photon flux of $7.3\cdot10^5~\mathrm{photons/s}$. For this photon flux the probability of having more than one photon per detector dead time is $0.25\%$. For the lower limit, we take the variance of the dark count noise. With the dark count noise $V_\mathrm{dark~counts}=(18\pm2)~\left[\mathrm{counts/s}\right]^2$, we calculate a shot-noise clearance of $(46.0\pm1.1)~\mathrm{dB}$~\cite{Jin2015, Bruynsteen2021}.

In addition to the shot-noise clearance, we calculated the common-mode rejection ratio (CMRR) from the calculated count rates of the SNSPDs at an LO photon flux of $\approx 560~\mathrm{kphotons/s}$. The CMRR characterizes the detector's capability to subtract efficiently amplitude fluctuations of the two beams incident on the two detectors after the beam splitter within the BHD. We calculate the CMRR as follows:
\begin{equation}
	\mathrm{CMRR}=10\log\left(\frac{\mathrm{CR}_\mathrm{1,dark~counts}-\mathrm{CR}_\mathrm{2, 50:50}}{\mathrm{CR}_\mathrm{1,50:50}-\mathrm{CR}_\mathrm{2, 50:50}}\right)[\mathrm{dB}] \,,
 \label{eq:CMRR}
\end{equation}
by measuring the ratio of the difference in photon-number count rates when one detector is blocked to the difference in photon-number count rates when both detectors are unblocked for a given input local-oscillator strength and vacuum at the other input port of the beam splitter. $\mathrm{CR}_\mathrm{1,2}$ is defined as the photon-number count rate of the detectors $1~\mathrm{and}~2$. The subscripts $\mathrm{dark~counts}$ and $\mathrm{50:50}$ refer to the dark count rate and the count rate in the balanced operation. With Eq.~(\ref{eq:CMRR}), we calculate a $\mathrm{CMRR}_\mathrm{opt.}=22.4~\mathrm{dB}$ for the polarization optimized balanced operation and $\mathrm{CMRR}_\mathrm{no~opt.}=14.3~\mathrm{dB}$ without polarization tuning. Due to the low LO photon flux, the relative influence of Poisson statistics in the photon number distribution is much larger than when using conventional photodiodes. To reduce the influence and increase the CMRR, one needs to go to higher photon fluxes, which would require faster detectors or SNSPD arrays, allowing for higher detection rates.

\section{Homodyne detection of a coherent state}
To demonstrate a homodyne experiment, we further investigated a weak coherent state and compared it to the measurement outcomes for the vacuum state. The weak coherent state and LO are derived from the same laser source using a fiber beam splitter ensuring mode matching between LO and signal state. Here, we consider an LO photon flux below $2~\mathrm{photons}$ within the coherence time of the laser. To investigate the quadratures of a weak signal~\cite{Vogel1993}, we can consider the nonlinear, normally order expectation values and variance of click-based quadrature operators~\cite{Sperling2015a}. This has been experimentally investigated in the frame of Stokes operators from a polarization-entangled light source~\cite{Prasannan2022}. To investigate the signal state characteristics with the SNSPDs, we calculate the nonlinear, normally ordered quadrature operator~\cite{Sperling2015a}
\begin{widetext}
    \begin{equation}
        \langle{:}\hat{X}^m(\varphi){:}\rangle=\sum^m_{j=0}\sum^N_{k_1=j}\sum^N_{k_2=m-j}(-1)^{m-j}N^m\frac{\binom{m}{j}\binom{k_1}{j}\binom{k_2}{m-j}}{\binom{N}{j}\binom{N}{m-j}}c_{k_1,k_2}
        \label{eq:Quadrature}
    \end{equation}
\end{widetext}
using the methods developed in~\cite{Sperling2015b}. The notation ${:}\ldots{:}$ indicates normal ordering of the creation and annihilation operators comprising the operator $\hat{X}$. In this equation, $c_{k_1,k_2}$ is the joint click statistics and describes the probability of measuring $k_1$ clicks in one detector and $k_2$ clicks in the other within the sampling time, and $N$ is the number of possible measurement outcomes. Equation~\ref{eq:Quadrature} is developed for BHD with arrays of on-off detectors in which light is uniformly distributed onto $N$ on/off detectors. Here, this is realized by a multiplexing into $N$ time bins whose size is determined by the ratio of sampling time and dead time. With the given sampling time of $2.5~\mathrm{\upmu s}$ and the dead time of $100~\mathrm{ns}$ from the SNSPDs, we have $N=26$. $m$ describes the moment of the nonlinear quadrature operator. While for the first moment ($m=1$), the expectation value of the nonlinear quadrature operator $\langle{:}\hat{X}(\varphi){:}\rangle$ is calculated from Eq.~(\ref{eq:Quadrature}) directly, the second moment is required to calculate the variance of the nonlinear quadrature operator according to~\cite{Sperling2015a},
\begin{equation}
    \langle{:}\left[\Delta\hat{X}(\varphi)\right]^2{:}\rangle=\langle{:}\hat{X^2}(\varphi){:}\rangle - \langle{:}\hat{X}(\varphi){:}\rangle^2,
    \label{eq:Variance}
\end{equation}
where $\langle{:}\hat{X^2}(\varphi){:}\rangle$ is the expectation value of the second moment of the nonlinear quadrature operator (i.e., found by setting $m=2$ in Eq.~(\ref{eq:Quadrature})). A common quadratic error propagation is used to determine uncertainties.
\begin{figure*}[ht!]
\centering\includegraphics[scale=1]{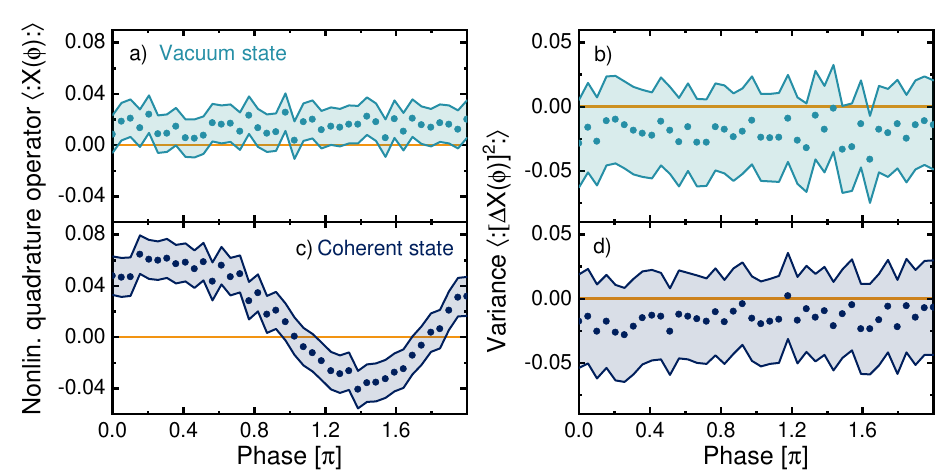}
\caption{Results for the mean of the nonlinear quadrature operator $\langle{:}\hat{X}(\varphi){:}\rangle$ and the variance $\langle{:}\left[\Delta\hat{X}(\varphi)\right]^2{:}\rangle$ are shown for different phases. They were calculated for the vacuum state (a) and (b) and for a weak coherent state with a photon flux of $\approx4.3~\mathrm{kphotons/s}$ (c) and (d). The optical phase difference between the LO and the two states is swept resulting in the oscillating behavior for the weak coherent signal state as expected. The vacuum state is phase-independent; therefore no oscillation is visible. The error is given by the three-standard-deviation environment. In orange, the expectation value for the vacuum state is highlighted. In case of the coherent state, we expect an oscillation around zero for $\langle{:}\hat{X}(\varphi){:}\rangle$ while the variance should be equal that of the vacuum state.}
\label{fig:5}
\end{figure*}

For the vacuum state, we expect a phase-independent behavior with the mean $\langle{:}\hat{X}(\varphi){:}\rangle_\mathrm{vac}=0$ and second moment $\langle{:}\hat{X^2}(\varphi){:}\rangle_\mathrm{vac}=0$. To measure this, we used a fiber-coupled phase modulator in the arm of the LO which was operated at a fixed photon flux of $\approx561~\mathrm{kphotons/s}$. We used a sample time of $2.5~\mathrm{\upmu s}$ as in the previous measurements to achieve the highest resolution for the quadrature measurement. From the joint click counting statistics at different phases, we calculate from Eq.~(\ref{eq:Quadrature}) the nonlinear quadrature operator for the vacuum as shown in Fig.~\ref{fig:5}~(a). The variance displayed in Fig.~\ref{fig:5}~(b) is calculated according to Eq.~(\ref{eq:Variance}).

For the mean $\langle{:}\hat{X}(\varphi){:}\rangle$ and variance $\langle{:}\left[\Delta\hat{X}(\varphi)\right]^2{:}\rangle$, the expected phase-independent behavior is visible. The offset of the mean from zero can be attributed to the imperfect CMRR due to the slight unequal splitting of the beam splitter and balancing the detector signals with the polarization control. Within a three-standard-deviation range, the normally ordered variances are consistent with the predicted value of zero for a non-squeezed vacuum state. Hence, our BHD together with the appropriate click-based theory functions as expected.

In addition, we investigated a weak coherent state. The coherent state is generated by splitting a laser beam at a 99:1 beam splitter: the strong port becomes the LO, and the weak port the coherent state (signal). The coherent state is further attenuated which gives a weak coherent state with $\approx4.3~\mathrm{kphotons/s}$. The fields of the brighter LO and the weak coherent state are then interfered on the balanced beam splitter.

From the joint click counting statistics, we calculate the mean $\langle{:}\hat{X}(\varphi){:}\rangle$ and variance $\langle{:}\left[\Delta\hat{X}(\varphi)\right]^2{:}\rangle$ as shown in Fig.~\ref{fig:5}~(c) and Fig.~\ref{fig:5}~(d). For the mean $\langle{:}\hat{X}(\varphi){:}\rangle$, a clear oscillation of the quadrature operator with the phase can be seen. With the phase modulator, we scan through $2\pi$ phases (compare Fig.~\ref{fig:5}~(c)). As previously mentioned for the vacuum state, a slight offset from zero is visible. The results for the variance $\langle{:}\left[\Delta\hat{X}(\varphi)\right]^2{:}\rangle$ match the results obtained for the vacuum state as expected.

\section{Conclusion}
We have demonstrated a continuous-wave low-noise balanced homodyne detector with superconducting nanowire single-photon detectors. We show that a homodyne detector with SNSPDs has a linear response to the input photon flux over a broad range of local oscillator photon fluxes. The range of LO photon fluxes is limited by the probability for multiphoton events within the dead time of the detector. By exploiting their low dark count rates, we measure a maximum shot-noise clearance of $(46.0\pm1.1)~\mathrm{dB}$. In addition, we have demonstrated a homodyne measurement on the vacuum state and a weak coherent state and calculated the nonlinear quadrature operator from the joint click counting statistics for different optical phases between the LO and the signal field.

Given the overall efficiency of $(77\pm7)\%$, such a detector could be used to measure up to $6.3~\mathrm{dB}$ squeezing which is well within the range allowed by the shot-noise clearance. Further improvements to overall system efficiency would increase this value. 
In order to extend the shot-noise clearance and operating speed even further, SNSPDs with lower dark count rates and fast recovery times can be used. Lower dark count rates will cause the flattening of the linear regime to shift to lower local oscillator photon fluxes, and faster recovery times will enable a higher sampling rates or smaller measurement times at the same sampling rate. The utilization of  SNSPDs arrays could potentially enable an even higher shot-noise clearance since the detectors can be operated at a higher input photon flux, without saturating.

In principle, integrating SNSPDs brings many advantages in the continuous variable regime, not least intrinsic phase stability. Furthermore, integrated SNSPDs have been shown with near-unity on-chip detection efficiency~\cite{Akhlaghi2015}. Together with our result, this can enable the development of high efficiency homodyne detectors utilizing single-photon-sensitive detectors.
\newpage
\begin{acknowledgments}
The authors thank Thomas Hummel, Michael Stefszky and Brian Smith for fruitful discussions. Partially funded by the European Union (ERC, QuESADILLA, 101042399). Views and opinions expressed are however those of the author(s) only and do not necessarily reflect those of the European Union or the European Research Council Executive Agency. Neither the European Union nor the granting authority can be held responsible for them. This work has received funding from the German Ministry of Education and Research within the PhoQuant project (grant number 13N16103) and ISOQC project (grant number 13N14911).
\end{acknowledgments}

\bibliography{library}

\end{document}